\documentstyle[aps,twocolumn,epsfig]{revtex}

\def\be{\begin{equation}}
\def\ee{\end{equation}}
\def\bea{\begin{eqnarray}}
\def\eea{\end{eqnarray}}
\def\bma{\begin{mathletters}}
\def\ema{\end{mathletters}}

\def\C{\hbox{$\mit I$\kern-.7em$\mit C$}}
\def\st{\mbox{ s. t. }}
\newcommand{\one}{\mbox{$1 \hspace{-1.0mm}  {\bf l}$}}
\def\bg{\begin{guess}}
\def\eg{\end{guess}}
\newcommand{\bra}[1]{\mbox{$\langle #1 |$}}
\newcommand{\ket}[1]{\mbox{$| #1 \rangle$}}
\newcommand{\kepsi}{\mbox{$|\psi^{< k} \rangle$}}
\newcommand{\brapsi}{\mbox{$\langle \psi^{< k}|$}}
\newcommand{\bracket}[2]{\mbox{$\langle {{#1}} \mathrel{ | {\vphantom
        {{#1} {#2}}} \kern-\nulldelimiterspace} {{#2}} \rangle$}}

\newcommand{\rem}[1]{}

\newcommand{\calp}{\mbox{$\cal P$}}

\begin{document}
\draft

\title{Quantum Correlations in Two-Fermion Systems}

\author{John Schliemann$^1$, J. Ignacio Cirac$^2$, Marek Ku\'s$^3$, Maciej
Lewenstein$^4$, and Daniel Loss$^5$}

\address{$^1$ Department of Physics, The University of Texas, Austin,
TX 78712\\
$^2$
Institut f\"ur Theoretische  Physik,
Universit\"at  Innsbruck, A-6020 Innsbruck, Austria\\
$^3$ Centre for Theoretical Physics, Polish Academy of Sciences, 02668 Warsaw,
Poland\\
$^4$
Institut f\"ur Theoretische  Physik,
Universit\"at  Hannover, 30167 Hannover, Germany\\
$^5$ Department of Physics and Astronomy, University of Basel, CH-4056
Basel, Switzerland}

\date{\today}

\maketitle

\begin{abstract}
We characterize and classify quantum correlations in two-fermion systems
having $2K$ single-particle states. For pure states we introduce the Slater 
decomposition and rank (in analogy to Schmidt decomposition and rank), 
i.e. we decompose the state into a combination of elementary Slater
determinants formed by pairs of mutually orthogonal single-particle states. 
Mixed states can be
characterized by their Slater number  which is the minimal Slater rank 
required to generate them. For $K=2$ we give a necessary and sufficient 
conditions for a state to have a Slater number one. We introduce a correlation
measure for mixed states which can be evaluated analytically for $K=2$.
For higher $K$, we 
provide a method of constructing and optimizing Slater number witnesses, 
i.e. operators that detect Slater numbers for some states. 
\end{abstract}

\narrowtext


\narrowtext

\section{introduction}

In recent years a lot of effort \cite{review,lewen002} in 
Quantum Information Theory (QIT) has been devoted to the
characterization of entanglement, which is one of the key features
of quantum mechanics\cite{rew2000}.  The resources
needed to  implement a particular protocol of quantum information
processing (see e.g.
\cite{ekert}) are closely linked to the entanglement properties of the
states used  in the protocol. In particular, entanglement lies at the heart of
quantum computing\cite{rew2000}. The most fundamental question with
regard to entanglement is: given a state of a multiparty system, is it
entangled or not (i.e. is it separable\cite{werner})? If the answer is
positive, then the next question is how strong the entanglement is. 
For pure states in bipartite systems the latter question can
be answered by looking at the Schmidt decomposition\cite{Peres}, the i.e.
decomposition of the vector in a product basis of the Hilbert space
with a minimal number of terms. For mixed states already the first
question is notoriously hard  to answer. 
There exist, however, many separability
criteria, such as the Peres-Horodecki criterion \cite{peres96,horo96}, and
more recent concepts such as entanglement witnesses and the 
corresponding ``entanglement revealing'' positive maps \cite{terhal,opti}.

While entanglement plays an essential role in quantum communication
between parties separated by macroscopic distances, the
characterization of quantum correlations at short distances is also an
open problem, which has received much less attention so far.
In this case the indistinguishable character of the particles
involved (electrons, photons,...) has to be taken into account. 
In his classic book, Peres \cite{Peres} discussed the entanglement 
in elementary states of indistinguishable particles.
These are symmetrized or antisymmetrized product states for bosons 
and fermions, respectively. It is easy to see that all such
states of two-fermion systems, and as well all such states 
formed by two non-collinear single-particle states 
in two-boson systems, are necessarily entangled in the usual sense.
However, in the case
of particles far apart from each other, this type of 
entanglement is not of physical relevance:   
{\sl ``No quantum prediction, referring to an atom located 
in our laboratory, is affected by the mere presence of similar atoms in remote
parts of the universe''}\cite{Peres}. 
This kind of entanglement between indistinguishable
particles being far apart from each other 
is not the subject of this paper. Our aim here is rather to classify and 
characterize the quantum correlations between indistinguishable 
particles (in our case fermions) at short distances.
We discuss below why this problem is relevant for 
quantum information processing in various physical systems. 

For indistinguishable particles  a pure
quantum state must be formulated in terms of Slater determinants or
Slater permanents for fermions and bosons, respectively.
Generically, a Slater determinant contains correlations due to the exchange 
statistics of the indistinguishable fermions. As the simplest possible
example consider a wavefunction of two (spinless) fermions,
\be
\Psi(\vec r_{1},\vec r_{2})=\frac{1}{\sqrt{2}}
\left[\phi(\vec r_{1})\chi(\vec r_{2})-\phi(\vec r_{2})\chi(\vec r_{1})
\right]
\label{fermwv}
\ee
with two orthonormalized single-particle wavefunctions 
$\phi(\vec r)$, $\chi(\vec r)$. Operator matrix elements between such 
single Slater
determinants contain terms due to the antisymmetrization of coordinates
(``exchange contributions'' in the language of Hartree-Fock theory).
However, if the moduli of $\phi(\vec r)$, $\chi(\vec r)$ have only
vanishingly small overlap, these exchange correlations will also tend
to zero for any physically meaningful operator. This situation is generically
realized if the supports of the single-particle wavefunctions are essentially
centered around locations being sufficiently apart from each other, or the
particles are separated by a sufficiently large energy barrier. 
In this case the
antisymmetrization present in Eq.~(\ref{fermwv}) has no physical effect.

Such observations clearly 
justify the treatment of indistinguishable particles separated by macroscopic
distances as effectively distinguishable objects. So far, research in 
Quantum Information Theory has concentrated on this case, where the exchange
statistics of particles forming quantum registers could be neglected, or was 
not specified at all. 

The situation is different if the particles constituting, say, qubits are
close together and possibly coupled in some computational process.
This the case for all proposals of quantum information
processing based on quantum dots technology 
\cite{divinloss,BLD:99,schlie}. Here qubits are
realized by the spins of electrons living in a system of quantum dots.
The electrons have the possibility of tunneling eventually from one dot to the
other with a probability which can be modified by varying external parameters
such as gate voltages and magnetic field. In such a situation the fermionic
statistics of electrons is clearly essential. 

Additional correlations in many-fermion-systems
arise if more than one Slater determinant  is involved, i.e. if
there is no single-particle basis such that a
given state of $N$ indistinguishable fermions can be
represented  as an elementary  Slater determinant (i.e. fully antisymmetric
combination of $N$ orthogonal single-particle states).
These correlations are  the analog of quantum entanglement in 
separated systems and are essential for
quantum information processing in non-separated systems.

As an example consider a ``swap'' process exchanging the spin states of 
electrons on coupled quantum dots by gating the tunneling amplitude between 
them \cite{BLD:99,schlie}. Before the gate is turned on, the two electrons in
the neighboring quantum dots are in a state represented by 
a simple Slater determinant, and can be regarded as distinguishable since they
are separated by a large energy barrier. When the barrier is lowered, 
more complex correlations between the electrons due to the dynamics arise. 
Interestingly, as shown in Refs. \cite{BLD:99,schlie}, 
during such a process the system must necessarily enter a highly
correlated  state that cannot be represented by a single Slater determinant. 
The final state of the gate operation, however, 
is, similarly as the initial 
one, essentially given by a single Slater determinant. 
Moreover, by adjusting the gating time
appropriately one can also perform a ``square root of a swap'' which turns
a single  Slater determinant into a ``maximally'' correlated state
in much the same way \cite{schlie}. In the end of such a process the electrons
can again be viewed as effectively distinguishable, but are in 
a maximally entangled state in the usual sense of 
distinguishable separated particles. In this sense the highly correlated
intermediate state can be viewed as a resource for the production of
entangled states.

We expect that similar scenarios apply to other schemes of 
quantum information processing that involve cold particles (bosons or fermions)
interacting at microscopic distances at which the quantum 
statistics becomes essential. For instance, it should be of relevance 
for quantum computing models employing ultracold atoms in optical lattices
\cite{cirac}, or ultracold atoms in arrays of optical microtraps \cite{birkl}.

It is the purpose of the present paper to analyse the above type of 
quantum correlations  between indistinguishable fermions in more
detail. However, to avoid confusion with the existing literature
we shall reserve in the following the term ``entanglement'' for separated
systems and characterize the analogous quantum correlation phenomenon in 
non-separated fermionic systems by the notions of Slater rank and Slater 
number to be defined below.

We are going to formulate analogies with the theory of entanglement,
and translate several very recent results \cite{opti,mapy,schmidt}
concerning standard systems of distinguishable parties 
(Alice$\ne$ Bob) to the case of indistinguishable fermions.
In general we will deal with a system of two fermions each of which 
live in a
$2K$-dimensional single-particle space.

The plan of the paper is as follows: In Section \ref{purestates}
we discuss pure states, and formulate
the analog of Schmidt decomposition and rank -- Slater decomposition and
rank.  We then discuss a simple operational criterion 
for the case of two electrons in two neighboring quantum dots ($K=2$)
to determine whether  a
given state is  of Slater rank 1. This criterion was first derived in Ref.
\cite{schlie}. In Section
\ref{mixedstates} we define the concept of Slater number for mixed
states. For $K=2$ we present necessary and suficient condition for a 
mixed state to have the Slater number 1. 
This is an analog of the Peres-Horodecki
criterion\cite{peres96,horo96} in the Wootters 
formulation \cite{Wootters}.
In Section IV we extend the results of Section III and define a 
Slater correlation
measure which is the analog of the entanglement formation
measure\cite{bennett}. This quantity can be 
calculated analytically for the case $K=2$, in analogy to the Wootters 
result\cite{Wootters}.
In Section V we turn to the case
$K>2$ and introduce Slater number witnesses of canonical form (defined in
analogy to entanglement\cite{terhal,mapy} and Schmidt
number\cite{terhal001,schmidt} witnesses). We construct examples of such
$k$-Slater witnesses, which provide necessary conditions  for a given state to
have the Slater number smaller than $k$; we also discuss   
optimization of Slater witnesses. Finally, we analyze   the
associated\cite{jamio} positive maps. 
We close by discussing
further analogies, but also differences, between entanglement in
separated systems of distinguishable particles 
as opposed to quantum correlations in non-separated systems 
of indistinguishable particles.


\section{Slater rank of pure states}
\label{purestates}

We consider two indistinguishable fermions each of which lives in the 
single-particle
Hilbert space  ${\cal C}^{2K}$. This situation is given, e.~g., in a system
of two electrons in $K$ neighboring quantum dots where only the orbital 
ground state of each dot is taken into account. Alternatively one may think 
of, say, two quantum dots with an appropriate number of
orbital states available for the two fermions.

The states (density matrices) in
such a system are positive self-adjoint operators acting on the
antisymmetric space ${\cal A}({\cal C}^{2K}\otimes{\cal C}^{2K})$.
Let us first consider pure states, i.e. projectors on a vector
$|\Psi\rangle\in{\cal A}({\cal C}^{2K}\otimes{\cal C}^{2K})$. Let
$f_a$, $f_a^{\dag}$, $a=1,\ldots,2K$,
denote the fermionic annihilation and creation 
operators of single-particle states forming an orthonormal basis in 
${\cal C}^{2K}$, and $|\Omega\rangle$
denotes the vacuum state. Each vector in the two-electron space can be
represented as
$|\Psi\rangle=\sum_{a,b}w_{ab}f_a^{\dag}f_b^{\dag}|\Omega\rangle$, where
$w_{ab}=-w_{ba}$ is an antisymmetric matrix. We have the following
generalization of the Theorem 4.3.15 from Ref. \cite{mehta}, which will allow
us to define the fermionic analog of the Schmidt decomposition:

\newtheorem{guess}{Lemma}
\bg 
{\rm For any antisymmetric $N\times N$ matrix $A\ne 0$ there exists a
unitary transformation $U'$ such that $A=U'ZU'^T$, 
where the matrix $Z$ has blocks on the diagonal,
\begin{equation}
Z={\rm diag}\left[
Z_0, Z_1,\ldots ,Z_M
\right] ,\quad Z_0=0,\quad Z_i=\left[
\begin{array}{cc}
0 & z_i \\
-z_i & 0
\end{array}
\right],\label{z}
\end{equation}
and $Z_0$  is a $(N-2M)\times(N-2M)$ null matrix.}
\eg

\noindent{\it Proof:} Let $A$ be a $N\times N$, complex, antisymmetric
matrix acting on ${\cal C}^N$, $A=-A^T$,
hence $A^{\dag}=-A^*$. Let us define: $B:=AA^*=-AA^{\dag}$.
$B$ is hermitian, $B=B^{\dag}$, hence diagonalizable by a unitary
transformation: $B=UDU^{\dag}$, $UU^{\dag}=\one$, $D$ - diagonal.
Now consider $C:=U^{\dag}AU^*$. It is easy to check that $C$ is
antisymmetric,
$
C^T =-C
$, and normal $
CC^{\dag} =C^{\dag}C$.
Let us decompose $C$ into its real and imaginary parts:
$C=F+iG;\quad F,G$ are real ${N\times N}$ matrices.
Since $C$ is antisymmetric, so are $F$ and $G$. Since $C$ is normal,
$F$ and $G$ commute. Thus $F$ and $G$ are real, antisymmetric,
commuting matrices. Hence they can be simultaneously brought to block
diagonal
forms by a real orthogonal transformation\cite{mehta}, $F=OF_{bd}O^T$,
$G=OG_{bd}O^T$,
$O$ is a ${N\times N}$ matrix, $OO^T =I$, where
\begin{eqnarray}
F_{bd}&=&{\rm diag}\left[
X_{0},X_1, \ldots, X_K
\right] ,\nonumber\\
 G_{bd}&=&{\rm diag}\left[
Y_0, Y_1 ,\ldots ,Y_L
\right],
\end{eqnarray}
and $X_0$, $Y_0$ are null matrices of some dimensions,
$X_0=0$, $Y_0=0$, whereas $X_i$, $Y_i$ are
standard antisymmetric $2\times 2$ blocks:
\begin{equation}
X_i=\left[
\begin{array}{cc}
0 & x_i \\
-x_i & 0
\end{array}
\right] ,\quad Y_i=\left[
\begin{array}{cc}
0 & y_i \\
-y_i & 0
\end{array}
\right].
\end{equation}
Thus $C=OZO^T$ where
$Z$ has the form (\ref{z})
and, finally $
A=UCU^T=UOZO^TU^T=(UO)Z(UO)^T=U^\prime ZU^{\prime T}$
with $U^\prime$ unitary. $\Box$

\bg
{\rm Every vector in the antisymmetric space  ${\cal A}({\cal
C}^{2K}\otimes{\cal C}^{2K})$ can be represented in an appropriately
chosen basis in ${\cal C}^{2K}$ in a form of the Slater decomposition 
\be
|\Psi\rangle=\frac{1}{\sqrt{\sum_{i=1}^K |z_i|^2}}
\sum_{i=1}^{K}z_i f_{a_{1}(i)}^{\dag}f_{a_{2}(i)}^{\dag}|\Omega\rangle,
\label{schmidtsum}
\ee
where the states $f_{a_{1}(i)}^{\dag}|\Omega\rangle$,
$f_{a_{2}(i)}^{\dag}|\Omega\rangle$, $i=1, \ldots, K$, 
form an orthonormal basis in
${\cal C}^{2K}$, i.e. each of these single-particle states occurs only 
in one
term in the summation (\ref{schmidtsum}).
The number of nonvanishing coefficients $z_i$ (i.e. the
number of elementary  Slater determinants required to construct
$|\Psi\rangle$) is called the Slater rank.}
\eg

\noindent{\it Proof:} Let
$|\Psi\rangle=\sum_{a,b}w_{ab}f_a^{\dag}f_b^{\dag}|\Omega\rangle$. Note
that the change of basis in ${\cal C}^{2K}$ corresponds to a unitary
transformation of fermionic operators, 
$f_a^{\dag}= \sum_{b}U_{ba}(f'_{b})^{\dag}$, which implies that 
in the new basis $w'=UwU^{T}$. From Lemma 1 we may choose $U$ such that
$w'$ will have the form (\ref{z}), which provides the Slater
decomposition. $\Box$

From the point of view of applications in quantum dot computers, it is
important to be able to distinguish states with Slater rank 1 (which can
be easily prepared and detected) from those that involve more than one 
elementary Slater determinant. In general, given $|\Psi\rangle$ in
some basis, in order to check the Slater rank, one has to perform the
Slater decompositon. As we know from Ref. \cite{schlie}, the situation is
simpler for the case $K=2$, where we have:

\bg{\rm [Ref. \cite{schlie}] 
A vector
$|\Psi\rangle=\sum_{a,b=1}^4w_{ab}f_a^{\dag}f_b^{\dag}|\Omega\rangle$ in
${\cal A}({\cal C}^{4}\otimes{\cal C}^{4})$ has the Slater rank 1 iff 
\be
\eta(|\Psi\rangle)=\left|\sum_{a,b,c,d}\epsilon^{abcd}w_{ab}
w_{cd}\right|=0,
\label{crit}
\ee
where $\epsilon^{abcd}$ denotes the totally antisymmetric tensor in
${\cal C}^{4}\otimes{\cal C}^{4}\otimes{\cal C}^{4}\otimes{\cal C}^{4}$.}
\label{lemasc} 
\eg

\noindent{\it Remark:} The quantity $\eta(|\Psi\rangle)$ can be constructed
from the dual state 
\be
|\tilde\Psi\rangle=\sum_{a,b}\tilde w_{ab} 
f^{\dag}_af^{\dag}_b|\Omega\rangle
\ee
defined by the dual matrix
\be
\tilde w_{ab}=\frac{1}{2}\sum_{c,d}\epsilon^{abcd}w^*_{cd}\,.
\ee
With these definitions we have
\be
\eta(|\Psi\rangle)=\left|\langle\tilde\Psi|\Psi\rangle\right|\,.
\ee
The proof of this Lemma was presented first in Ref. \cite{schlie}. 
An alternative proof can be given using Lemma 1 and observing that
\be
\det w=\left(\frac{1}{8}\langle\tilde\Psi|\Psi\rangle\right)^{2}\,,
\label{det}
\ee
where $w$ is the antisymmetric $4\times 4$-matrix defining $|\Psi\rangle$.

In the Appendix we list some further useful properties of 
$\eta(|\Psi\rangle)$ and the relation of the dualization operation to
an antiunitary implementation of particle-hole-transformation.
An interestring further question are possible generalizations of the above 
result to the case of $K$ fermions having  a single particle space 
${\cal C}^{2K}$.


\section{Slater number of mixed states}
\label{mixedstates}

Let us now generalize the concepts introduced above to the case of mixed
states. To this end, we define the Slater number of a mixed state, 
in analogy to the Schmidt number for the case of distinguishable
parties \cite{terhal001,schmidt} : 

\newtheorem{guess8}{Definition} 
\begin{guess8}
\rm  {Consider a  density matrix
$\rho$ of a two fermion  system,  and all its possible convex
decompositions in terms of pure states, i.e.
$\rho=\sum_i p_i \ket{\psi_i^{r_i}}\bra{\psi_i^{r_i}}$,
where $r_i$ denotes the Slater rank of $\ket{\psi^{r_i}_i}$; the Slater  number
of $\rho$, $k$, is defined as $k=\min\{r_{\max}\}$,
where ${r}_{\max}$ is the maximum Slater rank within a decomposition,
and the minimum is taken over all decompositions.}
\end{guess8}
In other words,
$k$ is the minimal Slater rank of the pure states that are needed in order
to construct $\rho$, and there is a construction of $\rho$ that 
uses pure states with Slater rank not exceeding $k$.

Many of the results concerning Schmidt numbers can be  
transferred directly to the Slater number.  
For instance, let us denote the whole space of density matrices in
${\cal A}({\cal C}^{2K}\otimes{\cal C}^{2K})$ by
$Sl_K$, and the set of density matrices that have Slater number $k$ or
less,  by $Sl_k$. $Sl_k$ is a convex compact subset of
$Sl_K$;   a state from $Sl_k$ will be called a state of
(Slater) class $k$. Sets of increasing Slater number are embedded into
each other, i.e. $Sl_1\subset Sl_2\subset...Sl_k ...\subset Sl_K$.
In particular, $Sl_1$ is the set of states
that can be written as a convex combination of elementary Slater
determinants; $Sl_2$ is the set of states of Slater number 2, 
i.e. those that require at least one pure state of Slater rank 2 for
their formation, etc. 

The determination of the Slater number of a given state is in general 
a very difficult task.
Similarly, however, as in the case of separability of mixed states of
two qubits (i.e. states in 
${\cal C}^{2}\otimes{\cal C}^{2}$), and one qubit--one qutrit (i.e.
states in ${\cal C}^{2}\otimes{\cal C}^{3}$) \cite{horo96}, the
situation is particularly simple in the case of small $K$. For $K=1$
there exists only one state (a singlet). For $K=2$
we will present below a necessary and sufficient condition for a given mixed 
state to have a Slater number of one. 
One should note, however, that in the considered case  of
fermionic states there exists no simple analogy of the partial
transposition, which is essential for the theory of entangled states. 
In fact, the Peres--Horodecki criterion\cite{peres96,horo96} in  $2\times
2$ and $2\times 3$ spaces says that a state is separable iff its partial
transpose is positive. It is known, however, that the
Peres-Horodecki criterion is equivalent to Wootters' result \cite{Wootters},
relating separability to a quantity called
concurrence, which is related to eigenvalues of a certain matrix. This
latter approach can be used to characterize fermionic states in ${\cal
A}({\cal C}^{4}\otimes{\cal C}^{4})$. We have the following Theorem:    

\newtheorem{guess2}{Theorem}
\begin{guess2}
{\rm Let the mixed state acting in ${\cal A}({\cal C}^{4}\otimes{\cal
C}^{4})$ have a spectral decomposition
$\rho=\sum_{i=1}^r|\Psi_i\rangle\langle\Psi_i|$, where $r$ is the rank of
$\rho$, and the eigenvectors $|\Psi_i\rangle$ belonging to nonzero
eigenvalues $\lambda_i$ 
are normalized as
$\langle\Psi_i|\Psi_j\rangle=\lambda_i\delta_{ij}$. 
Let $|\Psi_i\rangle=\sum_{a,b}w^i_{ab}
f^{\dag}_af^{\dag}_b|\Omega\rangle$ in some basis, and define the complex 
symmetric $r\times r$ matrix $C$ by
\be
C_{ij}=\sum_{abcd}\epsilon^{abcd}w^i_{ab}w^j_{cd},
\label{cma}
\ee
which   can be represented using a unitary matrix as 
$C=UC_dU^T$, with $C_d={\rm diag}[c_1,c_2,\ldots,c_r]$ diagonal and 
$|c_1|\ge|c_2|\ge \dots\ge |c_r|$. The state $\rho$ has Slater number 1
iff 
\be
|c_1|\le \sum_{i=2}^r |c_i|.
\ee}
\end{guess2}

\noindent{\it Proof:} Let us assume that a state 
$\rho$ acting in ${\cal A}({\cal C}^{4}\otimes{\cal
C}^{4})$ has Slater number 1, i.e. 
\be
\rho=\sum_{i=1}^r|\Psi_i\rangle\langle\Psi_i|=\sum^{r'}_{k=1}
|\phi_k\rangle\langle
\phi_k|,
\label{exp}
\ee
where all $\phi_k$ have Slater rank 1, whereas $r'$ can be an arbitrary
integer $\ge r$. But $|\phi_k\rangle$ can be represented as 
$|\phi_k\rangle=\sum_{i=1}^rU_{ki}|\Psi_i\rangle=
\sum_{i=1}^r\sum_{a,b}U_{ki}w^i_{ab}
f^{\dag}_af^{\dag}_b|\Omega\rangle$. From Lemma \ref{lemasc}
we obtain that for each $k$, $\eta(w'(k))=0$, where
$w'(k)_{ab}=\sum_{i=1}^rU_{ki}w^i_{ab}$. The matrices $U_{ki}$ must
therefore fulfill for every $k$ 
\be
\sum_{abcd}\sum_{i,j=1}^r\epsilon^{abcd}w^i_{ab}w^j_{cd}U_{ki}U_{kj}=
\sum_{i,j=1}^r C_{ij}U_{ki}U_{kj}=0.
\label{conc}
\ee
On the other hand, from Eq. (\ref{exp}) we obtain
\be
\sum^{r'}_{k=1} U_{ki}U^*_{kj}=\delta_{ij}.
\label{uni}
\ee
The Slater rank 1 is thus equivalent to the existence
of the $r'\times r$ martix $U_{ki}$ that fulfills Eqs. (\ref{conc}) and 
(\ref{uni}). It is convenient to represent the rows of the matrix
$U_{ki}$ as vectors $|R_k\rangle$ in an $r$ dimensional Hilbert space 
${\cal H}_{aux}$. Eqs.  (\ref{conc}) and  (\ref{uni}) then reduce to
$\sum_k^{r'}|R_k\rangle\langle R_k|=\one$,  and $\langle
R_k^*|C|R_k\rangle=0$ for all $k$. One can always change the basis in
${\cal H}_{aux}$, i.e. replace $|R_k\rangle\to U|R_k\rangle$. Such a
transformation does not affect  Eq. (\ref{uni}), and transforms
$C\to U^TCU$. Since $C$ is symmetric, $U$ can be choosen in such a way
that $U^TCU$ is diagonal, and Eq. (\ref{conc}) reads then
$\sum_{i=1}^r c_iU_{ki}^2=0$. In this new basis the construction of
$U_{ki}$ using the method of Wootters \cite{Wootters} can be carried over.
One can always assume that $c_1 U_{k1}^2$ is real and positive, by
chosing the phases of $|R_k\rangle$. Then, one observes that,
provided Eq. (\ref{conc}) is fulfilled, 
\be
0=\left|\sum_{i=1}^r c_iU_{ki}^2\right|\ge |c_1||U_{k1}^2|-\sum_{i=2}^r
|c_i||U_{ki}^2|.
\ee
Summing the above inequality over $k$ and using Eq. (\ref{uni}), we
obtain the necessary condition
\be
|c_1|\le \sum_{i=2}^r |c_i|.
\label{necc}
\ee
To show that it is also a sufficient condition, we take $r'=2$ if
$r=2$, 
$r'=4$ if $r=3,4$, $r'=8$ if $r=5,6$, and 
$U_{ki}=\pm 1_{ki}\exp(i\theta_i)/\sqrt{r'}$. The equations in Eq. (\ref{conc}) are then all
equivalent to 
\be
|c_1|=\sum_{i=2}^r c_i\exp(2i\theta_i),
\label{con}
\ee
 and the angles
$\theta_i$ can indeed be choosen to assure that Eq. (\ref{con}) is
fulfilled, provided the condition (\ref{necc}) holds. The $\pm 1_{ki}$
signs are designed in such a way that Eq. (\ref{uni}) is fulfilled. Thus
for
$r'=2$ we take $(++),(+-)$ for  $i=1,2$,   for $r'=4$ we take $(++++),
(++--),(+-+-), (+--+)$ for $i=1,\ldots,4$ (or any 3 of them for
$i=1,\ldots,3$), and finally for
$r'=8$,
$(++++++++), (++++----), (++--++--), (++----++), (+-+-+-+-),
(+-+--+-+)$. In the latter case we take again as many vectors as we
need, i.e. $i=1,\ldots,5\le r\le 6$. 

The above Theorem is an analog of the Peres-Horodecki-Wootters result for
two-fermion systems having a single-particle space of dimension
$2K\le 4$.  The
situation is much more complicated, when we go to
$K>2$; this is similar to the case of the separability problem in ${\cal
C}^M\otimes{\cal C}^N$ with $MN>6$. These issues are investigated in Section
\ref{witnesses}. In the following section, however, we shall
concentrate on the case $K=2$.
 

\section{Slater correlation measure}

The similarity of our approach to that of Wootters \cite{Wootters} 
can be pushed further, and in
particular allows us to define and calculate, for the case of $K=2$, the 
``Slater formation measure'' 
(in analogy to entanglement formation measure \cite{bennett}).

To this aim  we first consider a pure (normalized) state
$|\bar\psi\rangle=\sum_{a,b}w_{ab} f^{\dag}_af^{\dag}_b|\Omega\rangle$, and
define the {\em Slater correlation measure} of $|\bar\psi\rangle$ as in
Lemma~\ref{lemasc} (cf. Ref.~\cite{schlie}),
\be
\eta(|\bar\psi\rangle)=|\langle\tilde{\bar\psi}|\bar\psi\rangle|.
\label{meapur}
\ee
with $|\tilde{\bar\psi}\rangle$ being the dual of $|\bar\psi\rangle$.
Obviously, the notion of dual states, as well as the function 
$\eta(.)$ in Eq.~(\ref{meapur}), can be defined also for unnormalized 
states. In the following we will denote such unnormalized 
states just as states occurring in the previous sections, i.e. without the bar.

The measure (\ref{meapur}) has all desired properties \cite{bennett,plenio},
such that it vanishes iff $|\bar\psi\rangle$ has Slater rank 1, and it is 
invariant with respect to local bilateral unitary operations, or, 
in another words, with respect to changes of the basis in the single particle 
space.

Having defined the measure for the pure states, we can consider the 
following definition:
 
\begin{guess8}
\rm  {Consider a  density matrix $\rho$ acting in ${\cal A}({\cal
C}^{4}\otimes{\cal C}^{4})$,  and all its possible convex
decompositions in terms of pure states, i.e.
$\rho=\sum_i \ket{\psi_i}\bra{\psi_i}=
\sum_i p_i \ket{\bar\psi_i}\bra{\bar\psi_i}$,
where the unnormalized states $\ket{\psi_i}=\sqrt{p_i}\ket{\bar\psi_i}$; 
the Slater correlation measure of $\rho$, $Sl(\rho)$ 
is defined as 
$$Sl(\rho)={\rm{inf}}
\left\{\sum_i p_i\eta({|\bar\psi_i\rangle})\right\},$$
where the infimum is taken over all decompositions.}
\end{guess8}
In other words,
$Sl(\rho)$ is the minimal amount of Slater 
correlations of the pure states that are needed in order
to construct $\rho$, and there is a construction of $\rho$ that 
uses pure states with ``averaged'' Slater correlation $Sl(\rho)$.

Note that $\sum_i p_i\eta({|\bar\psi_i\rangle})=\sum_i
\eta({|\psi_i\rangle})$. As we shall see below,
the measure $Sl(\rho)$ can be related directly to
the matrix $C_{ij}$ in Eq. (\ref{cma}), and to its ``concurrence''.  
It is invariant not only with respect to local
bilateral unitary operations, but it also cannot increase under local
bilateral operations. These are trace preserving maps of the form
$\rho\to M(\rho)=\sum_j A_j\otimes A_j \rho A^{\dag}_j\otimes
A^{\dag}_j$, where  each $A_j$ acts in ${\cal C}^4$, and $ \sum_j
A^{\dag}_jA_j
\otimes  A^{\dag}_jA_j =\one$. Such transformations correspond 
to mixtures of density matrices obtained after 
nonunitary changes of the basis in the single-particle space.
It is easy to see that 
$$Sl(M(\rho))=\left(\sum_j |{\rm \det}A_j|Sl(\rho)\right)\le 
Sl(\rho)$$. 

We have the following theorem:
\begin{guess2}
For any $\rho$ acting in ${\cal A}({\cal
C}^{4}\otimes{\cal C}^{4})$
\be
Sl(\rho)=|c_1|-\sum_{i=2}^{r'}|c_i|,
\ee
where $c_i$ are the diagonal elements of $C$ (Eq. (\ref{cma})) in the basis 
that diagonalizes it.
\end{guess2}

\noindent{\it Proof:} The proof is essentially the same as the one in the 
previous section. Let us consider an arbitrary expansion of a given
density matrix, 
$\rho=\sum_{k=1}^{r'}|\phi_k\rangle
\langle \phi_k|$, where $|\phi_k\rangle=\sum_{j=1}^rU_{kj}|\Psi_j\rangle$.
Here  
$|\Psi_j\rangle$ denote the usual ``subnormalized'' eigenvectors of $\rho$ with
$\langle\Psi_j|\Psi_j\rangle$ being equal to the $j$-th nonzero
eigenvalue of $\rho$ \cite{Wootters}.
It is easy to see that
\be
Sl(|\phi_k\rangle\langle
\phi_k|)=\left|\sum_{i,j=1}^r C_{ij}U_{ki}U_{kj}\right|,
\ee
and $\sum_{k=1}^{r'}U^*_{ki}U_{kj}=\delta_{ij}$. By changing the basis to 
the one in which $C$ is diagonal we get (after choosing the phases of
$U_{k1}$  such that 
$c_1U^2_{k1}$ are real and positive):
\be
\sum_{k=1}^{r'}Sl(|\phi_k\rangle\langle \phi_k|)=\sum_{k=1}^{r'}
\left|\sum_{j}c_{j}U^2_{kj}\right|\ge |c_1|-\sum_{i=2}^{r'}|c_i|.
\ee
This inequality becomes an equality when we use the same construction of $U_{kj}$
as in previous section, namely $U_{kj}=
\pm 1_{kj}\exp(i\theta_j)/\sqrt{r'}$, with 
$\theta_j$ selected in such a way that (independently of $k$),
\be
\left|\sum_{j}c_{j}U^2_{kj}\right|=\frac{1}{r'}
\left(|c_1|-\sum_{i=2}^{r'}|c_i|\right).
\ee
$\Box$

The above construction provides, to our knowledge, a rare example of an analog
of the entanglement formation measure that can be evaluated analytically. 
Obviously, since we have 
introduced the concept of Slater coefficients, we may define other Slater 
correlations measures for pure states 
in terms of appropriately designed 
convex functions of the Slater coefficients 
(in analogy to entanglement monotones 
\cite{Vidal}). For $K=2$, and most probably 
only for $K=2$, all those measures are equivalent and the corresponding 
induced measures for mixed states can be calculated analytically.  


\section{Slater witnesses}
\label{witnesses}

We now investigate fermion systems with single-particle Hilbert
spaces of dimension $2K>4$. In this case, a full and explicit 
characterization of pure and mixed state 
quantum correlations, such as given above for the two-fermion system with
$K=2$, is apparently not possible.
Therefore one has to
formulate other methods to investigate the Slater number of a given state.
We can, however, follow here the lines of the papers that we have
written on entanglement witnesses\cite{opti,mapy}, and Schmidt number
witnesses\cite{schmidt}.

In order to determine the Slater number of a density matrix
$\rho$ we note that due to the fact that the sets $Sl_k$ are convex and
compact,  any  density matrix of class $k$  can be decomposed into
a  convex combination of a density matrix of class $k-1$,  and  a
remainder $\delta$\cite{Le98}:

\newtheorem{guess3}{Proposition}
\begin{guess3}
\rm { Any state of class $k$, $\rho_k$,  can be
written as a convex combination of a density matrix of class $k-1$ and
a so-called $k-$edge state $\delta$:
\be
\rho_k  =(1-p) \rho_{k-1} + p \delta, \;\;\;\, 1\ge p> 0 ,
\label{decom}
\ee
where the edge state $\delta$ has Slater  number $\ge k$.}
\end{guess3}

The decomposition (\ref{decom}) is obtained by
subtracting projectors onto pure states of Slater rank smaller than
$k$, $P=\ket{\psi^{<k}}\bra{\psi^{<k}}$
such that $\rho_k-\lambda P\ge 0$. Here $\ket{\psi^{<k}}$ stands for 
pure states of Slater rank $r <k$.
Denoting by $K(\rho)$, $R(\rho)$, and
$r(\rho)$ the kernel, range, and rank of $\rho$, respectively,   
we observe that $\rho'\propto \rho-\lambda {\kepsi}{\brapsi}$  
is non negative iff ${\kepsi}\in R(\rho)$ 
and $\lambda\le {\brapsi}\rho^{-1}{\kepsi}^{-1}$ (see \cite{Le98}).
The idea behind this decomposition is that the edge state $\delta$
which has generically lower rank contains all the information
concerning the Slater number $k$ of
the density matrix $\rho_k$. 

As in the case of Schmidt number, there 
is an optimal decomposition of the form (\ref{decom}) with $p$ minimal.
Alternatively, restricting ourselves to decompositions
$\rho_k=\sum_i p_i \ket{\psi_i^{r_i}}\bra{\psi_i^{r_i}}$
with all $r_i\le k$, we can always find a decomposition
of the form (\ref{decom}) with  $\delta\in Sl_k$. We define below more
precisely what an edge state is.

\begin{guess8}
\rm {A $k$-edge state $\delta$
is a state such that
$\delta-\epsilon {\kepsi}{\brapsi}$ is not positive,
for any $\epsilon>0$ and $\ket{\psi^{<k}}$.} 
\end{guess8}

\newtheorem{guess4}{Criterion}
\begin{guess4} \rm{A mixed state
$\delta$ is a $k$-edge state iff there exists no
 $\kepsi$ such that  $\kepsi\in R(\delta)$.}
\end{guess4}

Now we are in the position of defining 
 a $k$-Slater witness ($k$-SlW, $k\ge 2$):  
\begin{guess8}
\rm{ A hermitian operator $W$ is a Slater witness (SlW) of class $k$ 
iff Tr$(W\sigma)\ge 0$ for all $\sigma\in Sl_{k-1}$, and there exists 
at least one $\rho \in Sl_k$ such that Tr$(W\rho)< 0$.}
\end{guess8}
It is straightforward to see that every SlW that detects $\rho$
given by (\ref{decom}) also detects the edge state $\delta$,
since if Tr$(W\rho)<0$  then necessarily Tr$(W\delta)<0$, too. 
Thus, the knowledge of  all SlW's of $k$-edge states fully characterizes 
all $\rho\in Sl_{k}$. Below, we show how to
construct for any edge state a SlW which detects it.
Most of the technical proofs used to construct and optimize
Slater witnesses are very similar to those presented in 
Ref.\cite{opti} for entanglement witnesses.

All the operators we consider below act in ${\cal A}({\cal
C}^{2K}\otimes{\cal C}^{2K})$. Let $\delta$ be a $k$-edge state,
$C$ an arbitrary positive operator such that ${\rm Tr}(\delta
C)>0$, and $P$ a positive operator whose range fulfills
$R(P)=K(\delta)$. We define
$\label{epsilon1} 
\epsilon\equiv  \inf_{\kepsi}{\brapsi}P{\kepsi}$ and 
$ c\equiv \sup {\bra{\psi}}C{\ket{\psi}}$. 
Note that $c>0$ by construction and $\epsilon > 0$, 
because $R(P)= K(\delta)$ and 
therefore, since $R(\delta)$ does not contain any ${\kepsi}$ 
by the definition of edge state, $K(P)$ cannot contain any  
${\kepsi}$ either. This implies:
\bg
{\rm
Given a $k$-edge state $\delta$, then
\be
W = P-{\epsilon \over c}C
\ee
is a $k$-SlW which detects $\delta$.} 
\eg
The simplest choice of $P$ and $C$ consists in taking 
projections onto $K(\delta)$ and the identity
operator on the asymmetric space $\one_a$, respectively. As we will see
below, this choice provides us with a canonical form of a $k$-SlW. 

\begin{guess3} 
{\rm
Any Slater witness can be written it the {\it canonical} form:
\be
W=\tilde{W}-\epsilon \one_a\ ,
\ee
such that $R(\tilde W)= K(\delta)$, where $\delta$ is a
$k$-edge state and $0<\epsilon\le {\rm inf}_{|\psi\rangle\in
S_{k-1}}{\bra{\psi}}\tilde{W}{\ket{\psi}}$}.
\end{guess3}
\noindent {\it Proof:} Assume $W$ is an arbitrary $k$-SlW such
that Tr$(W\sigma)\ge 0$ for
all $\sigma\in Sl_{k-1}$, and there $\exists$ at least one $\rho$ such
that Tr$(W\rho)<0$.
$W$ has at least one negative
eigenvalue. Construct $W+\epsilon \one_a =\tilde{W}$, such that  $\tilde{W}$
is a positive operator on ${\cal A}({\cal
C}^{2K}\otimes{\cal C}^{2K})$ but does not have a full rank, 
$K({\tilde{W}})\neq\emptyset$ (by continuity this construction 
is always possible). But 
${\brapsi} \tilde{W}{\kepsi} \ge \epsilon >0$ since $W$ is a $k$-SlW,
{\it ergo} no ${\kepsi}\in K(\tilde{W})$.$\Box$\\

\begin{guess8}
{\rm A $k$-Slater witness $W$ is {\it tangent} to $Sl_{k-1}$ at $\rho$ 
if $\exists$ a state $\rho\in  Sl_{k-1}$ such that 
Tr$(W\rho)=0$}.
\end{guess8}

\newtheorem{guess5}{Observation}
\begin{guess5}
\rm{The state $\rho$ is of Slater class $k-1$ iff
for all $k$-SW's tangent to $Sl_{k-1}$, Tr$(W\rho)\ge 0$.}
\end{guess5}

\noindent{\it Proof} (See \cite{opti}): (only if) Suppose that $\rho$
is of class $k$. From the Hahn-Banach theorem it follows that
there $exists$ a $k$-SlW, $W$,
that detects it. We can subtract 
$\epsilon \one_a$ from $W$, making $W-\epsilon \one_a$ tangent to 
$Sl_{k-1}$ at some $\sigma$, but then Tr$(\rho(W-\epsilon\one))<0$.$\Box$

\subsection{Optimal Slater witnesses}

We will now discuss the optimization of a Slater witness.
As proposed in \cite{opti} and \cite{schmidt} an entanglement
witness (Schmidt witness) W is optimal  if there exists no other witness that
detects more states than it.  The same definition can be applied to
Slater witnesses.  We say that a $k-$Slater witness $W_2$ is  finer  
than   a $k-$Slater witness $W_1$, if 
$W_2$ detects more states than $W_1$. Analogously, we define a
$k-$Slater witness  $W$ to be optimal when 
there exists no finer witness than itself. 
Let us define the set of ${\kepsi}$ 
pure states of Slater rank $k-1$ 
for which the expectation value of the
$k$-Slater witness $W$ vanishes:
\be
T_{W}=\{ \kepsi \st \brapsi W \kepsi=0 \}\ ,
\ee
i.e. the set of pure tangent states of Slater rank $<k$.
$W$ is an optimal $k$-SlW iff $W-\epsilon P$ is not a $k$-SlW,
for any positive operator $P$. If the set
$T_{W}$ spans the whole Hilbert space ${\cal A}({\cal
C}^{2K}\otimes{\cal C}^{2K})$, then $W$ is an optimal $k$-SlW.
If $T_{W}$ does not span ${\cal A}({\cal
C}^{2K}\otimes{\cal C}^{2K})$, 
then we can optimize the witness by subtracting from 
it a positive operator
$P$, such that $PT_{W}=0$. For example,  
for Slater witnesses of class 2 this is possible provided that 
$\inf_{|e\rangle \in{\cal
C}^{2K}}[P_{e}^{-1/2}W_{e}P_{e}^{-1/2}]_{\rm min}>0$. 
Here for any $X$ acting on ${\cal A}({\cal C}^{2K}\otimes{\cal C}^{2K})$
we define
\begin{eqnarray}
X_{e} & = & \Big[\langle e,.|X|e,.\rangle 
-\langle e,.|X|.,e\rangle\nonumber\\
& - & \langle ., e|X|e,.\rangle + \langle .,e|X|.,e\rangle\Big]\,, 
\end{eqnarray}
as an operator acting in  ${\cal C}^{2K}$, and $[X]_{\rm
min}$ denotes its minimal eigenvalue (see\cite{opti}).
An example of an optimal witness of Slater  number $k$ 
in ${\cal A}({\cal
C}^{2K}\otimes{\cal C}^{2K})$ is given by
\be
W=\one_a-\frac{K}{k-1}\calp \;,
\label{example}
\ee
where $\calp$ is a projector onto a ``maximally correlated state'',
$|\Psi\rangle=\frac{1}{\sqrt{K}}
\sum_{i=1}^{K}f_{a_{1}(i)}^{\dag}f_{a_{2}(i)}^{\dag}|\Omega\rangle$
(cf. Eq.~\ref{schmidtsum})
The reader can easily check that the above witness operator has mean value zero
in the states  $f_{a_{1}(i)}^{\dag}f_{a_{2}(i)}^{\dag}|\Omega\rangle$ for 
$i=1,2$, but also
for all states of the form $g_{1}^{\dag}g_{2}^{\dag}|\Omega\rangle$ where
\begin{eqnarray}
g_1^{\dag} & = & f_{a_{1}(1)}^{\dag}e^{i\varphi_{11}}
+f_{a_{1}(2)}^{\dag}e^{i\varphi_{12}}\nonumber\\
 & & +f_{a_{2}(1)}^{\dag}e^{i\varphi_{21}}
+f_{a_{2}(2)}^{\dag}e^{i\varphi_{22}},\\
g_2^{\dag} & = & -f_{a_{1}(1)}^{\dag}e^{-i\varphi_{12}}
+f_{a_{1}(2)}^{\dag}e^{-i\varphi_{11}}\nonumber\\
 & & -f_{a_{2}(1)}^{\dag}e^{-i\varphi_{22}}
+f_{a_{2}(2)}^{\dag}e^{-i\varphi_{21}},
\end{eqnarray}
for arbitrary $\varphi_{ij}$, $i,j=1,2$. The set $T_W$ spans  in this
case the whole Hilbert space ${\cal A}({\cal
C}^{2K}\otimes{\cal C}^{2K})$, {\it ergo} $W$ is optimal.

\subsection{Slater witnesses and positive maps}

It is interesting to 
consider linear maps associated with Slater witnesses via the Jamio\l
kowski isomorphism\cite{jamio}. Such maps employ $W$ acting in ${\cal
H}_A\otimes{\cal H_B}={\cal C}^{2K}\otimes{\cal C}^{2K}$, and transform a
state $\rho$ acting in
${\cal H}_A\otimes{\cal H_C}={\cal C}^{2K}\otimes{\cal C}^{2K}$ into
another state acting in
${\cal H}_B\otimes{\cal H_C}={\cal C}^{2K}\otimes{\cal C}^{2K}$,
$M(\rho)={\rm Tr}_A(W\rho^T_A)$. Obviously, such maps are positive on
separable states: When $\rho$ is separable, then for any $|\Psi\rangle\in
{\cal H}_B\otimes{\cal H_C}$, the mean value of
$\langle\Psi|M(\rho)|\Psi\rangle$, becomes  a convex sum of mean values 
of $W$ in some product states $|e,f\rangle\in{\cal
H}_A\otimes{\cal H_B}$. Since $W$ acts in fact in the antisymmetric space,
we can antisymmetrize these states, i.e. $|e,f\rangle\to (|e,f\rangle-|f,e\rangle)$.
Such antisymmetric states have, however, Slater rank 1, and all SlW of
class $k\ge 2$ have thus positive  mean value in those states. This class
of positive maps is quite different from the ones considered in Refs.
\cite{opti,mapy}; they provide thus an interesting
class of necessary separability conditions. 
The map associated with the witness (\ref{example}) is, however, 
decomposable, i.e. it is a sum of a completely positive map and another 
completely positive map composed with transposition.  
This follows from the fact that 
the witness operator has a positive partial transpose, i.e. it  can 
be presented as a partial transpose of a positive operator. 


\section{Conclusions and Outlook}
\label{concl}

Summarizing, we have presented a general characterization of 
quantum correlated states in two-fermion systems with a $2K$-dimensional
single-particle space. This goal has been achieved by introducing the
concepts of Slater deomposition and rank for pure states, and Slater
number for mixed states.
In particular, for the important case $K=2$ 
the  quantum correlations in mixed states can be characterized completely
in analogy to Wootters' result for separated qubits \cite{Wootters},
and using the findings of Ref.\cite{schlie} for pure states.
Similarly to the case of separated systems, the situation for $K>2$ is more
complicated. Therefore, we have also introduced   witnesses of
Slater number $k$, and presented the methods of optimizing them. 

Possible directions for future work include generalizations of the present
results to more than two fermions, and the development of an analogous theory
for indistinguishable bosons. For this purpose a lot of the concepts
developed so far are expected to be useful there as well. However, there are
certainly also fundamental differences between quantum correlations
in bosonic and fermionic systems. As an example consider the
notion of unextendible product bases introduced recently for separated
systems
\cite{unex}. These are sets of product states spanning a subspace of the
Hilbert space whose orthogonal complement does not contain any product
states. All such unextendible product bases constructed so far involve
product states of the form $|\psi\rangle\otimes|\chi\rangle$ with 
$|\psi\rangle$ and $|\chi\rangle$ being non-orthogonal. In the  
analogous fermionic state non-orthogonal contributions are obviously 
cancelled out by antisymmetrization, unlike the bosonic case. In fact, 
all explicit constructions of unextendible product bases 
known so far \cite{unex} can be taken over directly to bosonic systems to give
``unextendible Slater permanent bases''. These are sets of 
symmetrized product states spanning a subspace of the symmetrized Hilbert 
space, whose orthogonal complement does not contain any  such states.

\acknowledgements{We thank Anna Sanpera and W.~K. Wootters
for useful discussions, and
Allan H. MacDonald for helpful comments and a critical reading of the 
manuscript. J.~S. was supported by the Deutsche 
Forschungsgemeinschaft under Grant No. SCHL 539/1-1.
M.~K. was supported by Polish KBN Grant No. 2 P03B 072 19.
M.~L. acknowledges support by the Deutsche 
Forschungsgemeinschaft via SFB 407, Schwerpunkt
``Quanteninformationsverarbeitung'', and the Projekt 436 POL 133/86/0.
D.~L. acknowledges partial support from the Swiss National Science Foundation.


\appendix
\section{}
\label{appendix1}
We now list further properties of the correlation measure $\eta$ for
pure states 
$|\Psi\rangle=\sum_{a,b=1}^4w_{ab}f_a^{\dag}f_b^{\dag}|\Omega\rangle$
of two fermions in a four-dimensional single-particle space \cite{schlie},
and add some further remarks.

The matrix $w$ transforms under a unitary transformation of the
one-particle space,
\be
f^{+}_{a}\mapsto{\cal U}f^{+}_{a}{\cal U}^{+}=\sum_{b}U_{ba}f^{+}_{b}\,,
\ee
as 
\be
w\mapsto UwU^{T}\,,
\ee
where $U^{T}$ is the transpose (not the adjoint) of $U$.
Under such a transformation, 
$|\Psi\rangle\mapsto|\Phi\rangle={\cal U}|\Psi\rangle$,
 scalar products of the form
$\langle\tilde\Psi_{1}|\Psi_{2}\rangle$ remain unchanged up to a phase,
\be
\langle\tilde\Phi_{1}|\Phi_{2}\rangle
=\det U\langle\tilde\Psi_{1}|\Psi_{2}\rangle\,.
\label{scalprod}
\ee
Therefore, in particular, $\eta(|\Psi\rangle)$ is invariant under 
arbitrary single-particle transformations. 

The dualisation of a state $|\Psi\rangle$ can be identified as a
particle-hole-transformation,
\be
{\cal U}_{p-h}f^{+}_{a}{\cal U}^{+}_{p-h}=f_{a}\quad,\quad
{\cal U}_{p-h}|\Omega\rangle
=f^{+}_{1}f^{+}_{2}f^{+}_{3}f^{+}_{4}|\Omega\rangle\,,
\label{ph1}
\ee
along with a complex conjugation. In fact, the operator of dualization
$\cal D$, 
$|\Psi\rangle\mapsto|\tilde\Psi\rangle={\cal D}|\Psi\rangle$, can be written as
\be
{\cal D}=-{\cal U}_{p-h}{\cal K}\,,
\ee
where $\cal K$ is the usual operator of complex conjugation which acts on
a general state vector as
\be
{\cal K}\left(a|\alpha\rangle+b|\beta\rangle\right)
=a^{\ast}{\cal K}|\alpha\rangle+b^{\ast}{\cal K}|\beta\rangle\,.
\ee 
Its action on the 
single-particle basis states and the fermionic vacuum is given by
\be
{\cal K}f^{+}_{a}{\cal K}=f^{+}_{a}\quad,\quad
{\cal K}f_{a}{\cal K}=f_{a}\quad,\quad{\cal K}|\Omega\rangle=
|\Omega\rangle\,.
\label{defK}
\ee 
The relations (\ref{defK})
are to be seen as a part of the definition of ${\cal K}$ and
refer explicitly to a certain single-particle basis
defined by the operators $ f_{a}$, $ f^{+}_{a}$. However, switching to
a different complex conjugation operator ${\cal K}^{\prime}$, fulfilling
the relations (\ref{defK}) in a different basis, has only trivial effects
without any physical significance. In particular, as one can see from
the properties given above, the correlation measure
$\eta(|\Psi\rangle)=|\langle\tilde\Psi|\Psi\rangle|$,
$|\tilde\Psi\rangle={\cal D}|\Psi\rangle$, remains invariant
under such an operation.

Eq.~(\ref{scalprod}) implies that $\cal D$ is unchanged by unitary 
single-particle operations,
\be
{\cal U}{\cal D}{\cal U}^{+}={\cal D}\quad\Leftrightarrow\quad
\left[{\cal U},{\cal D}\right]=0\,
\ee
which can also be expressed as
\be
{\cal U}{\cal U}_{p-h}{\cal U}^{T}={\cal U}_{p-h}
\ee
for any unitary single-particle transformation ${\cal U}$.

The dualisation operator $\cal D$ is the antiunitary implementation
of the particle-hole-transformation. We note that the complex 
conjugation involved there is necessary for $\cal D$ being compatible
with single-particle transformations $\cal U$,
\begin{eqnarray}
{\cal D}{\cal U}f^{+}_{a}{\cal U}^{+}{\cal D}^{-1} & = &
\sum_{b}U^{\ast}_{ba}f_{b}\nonumber\\
 & = & {\cal U}{\cal D}f^{+}_{a}{\cal D}^{-1}{\cal U}^{+}\,.
\end{eqnarray}
If the complex conjugation would be left out, $\cal U$ and $\cal D$ would
not commute.

The relation of the correlation measure $\eta$ to an antiunitary operator
is similar to Wootters' construction for a separate system of two qubits
\cite{Wootters}.
The correlation measure there (``concurrence'') relies on the time
inversion operation. The operator of time inversion in the two-qubit
system is invariant under local unitary transformations in each qubit space.
This property is similar to the invariance of the dualisation operator under
unitary transformations in the single-particle space.

\end{document}